# Structural origin of gap states in semicrystalline polymers and the implications for charge transport

Jonathan Rivnay[1], Rodrigo Noriega[2], John E. Northrup[3], R. Joseph Kline[4], Michael F. Toney[5], Alberto Salleo[1]
1. Materials Science and Engineering, Stanford University, Stanford, CA
2. Applied Physics, Stanford University, Stanford, CA
3. Palo Alto Research Center, Palo Alto, CA
4. National Institute of Standards and Technology, Gaithersburg, MD
5. Stanford Synchrotron Radiation Lightsource, Menlo Park, CA

ABSTRACT:
*We quantify the degree of disorder in the π-π stacking direction of crystallites of a high performing semicrystalline semiconducting polymer with advanced X-ray lineshape analysis. Using first principles calculations, we obtain the density of states of a system of π-π stacked polymer chains with increasing amounts of paracrystalline disorder. We find that for an aligned film of PBTTT the paracrystalline disorder is 7.3%. This type of disorder induces a tail of trap states with a breadth of ~100 meV as determined through calculation. This finding agrees with previous device modeling and provides physical justification for the mobility edge model.*

Increased basic understanding of charge transport in polymeric semiconductors has been instrumental in the development of commercially viable active materials for lighting, solar cells, sensing and logic circuits.[1] Such developments have occurred in spite of the morphological and microstructural complexity associated with these materials, which makes it challenging to paint a simple picture of their limitations.[2] As a result of structural heterogeneity at several length scales, it has been difficult to relate structure to charge transport using quantitative and predictive models, whose results can successively be used for the design of new materials. Hence, linking microstructure to electrical properties still stands as a fundamental challenge to our understanding of how organic semiconductors function.

Charge transport in organic solids is generally understood in two limiting cases. Transport in highly ordered crystals is often thought of as being band-like,[3] while in perfectly disordered materials, charges are localized at individual sites and move by a thermally activated hopping process.[4] Recent studies raise the possibility of charge localization due to dynamic disorder caused by thermal fluctuations in the intermolecular distances, a consequence of the weak van der Waals bonding.[5] The mechanism for charge transport is determined by the type of electronic states involved, whether they are localized or delocalized. To describe systems where elements of both order and disorder are present, a model that includes both localized and delocalized states should be used. One such model is the mobility edge (ME) model, in which a band of delocalized states is separated from a tail of localized wavefunctions that act as electronic traps.[6] The energy that demarcates the delocalized mobile states from the traps is known as the mobility edge. Transport through such a system is dictated by the position of the Fermi level with respect to the mobility edge –which determines the population of the mobile states– and the average mean free path of a charge carrier in a mobile state. The key parameters of this model are the mobility of the delocalized states $\mu_o$, the density of trap states $N_t$, and the breadth of the energy distribution of such trap states, $E_0$. The number of mobile states is considered large so that the electronic properties of the material are determined by the tail of the density of states (DOS).

Given its success in relating structural defects (i.e. grain-boundaries) to electronic defects in polycrystalline silicon,[7] it is reasonable to expect that the ME model may provide similar insights into those organic materials that exhibit a high degree of order. Indeed, the same model was more recently used to describe transport in polycrystalline small molecule [8, 9] and semicrystalline polymer semiconductors.[10-12] The existence of electronic trap states in the gap of organic semiconductors is widely observed in organic thin film transistors, resulting in a broad sub-threshold region. It has been shown that weak static disorder gives rise to exponential Urbach tails in organic solids, similarly to inorganic materials.[13] In some cases, a structural origin of these states has been found. For instance, gap states have been postulated to exist in pentacene films due to the sliding of single molecules along their axis.[14] In the vast majority of cases, however, the structural origin of electronic trap states is still completely unknown. This is especially the case in polymers, where detailed structural characterization is exceptionally difficult.

Semicrystalline polymers exhibit disordered amorphous regions in coexistence with crystalline regions.[15] Trap states are often assumed to appear due to the most disordered regions of the film, such as grain-boundaries, while, in this respect, little attention is paid to structural defects in the most ordered parts of the material. Disorder in the crystalline regions of the film, however, can occur as thermal -or non-cumulative- fluctuations, as well as in the form of cumulative distortions such as paracrystallinity, which is known to be prevalent in polymeric crystallites.[16] Paracrystalline disorder is described as a statistical fluctuation of individual lattice spacings.[17] Consider two adjacent unit cells in a column along a specific [hkl] direction. These two points would have an ideal separation of $d_{hkl}$ (where $d_{hkl}$ is the average lattice spacing) and an actual separation of $d'_{hkl}$. A paracrystallinity parameter[16] $g$ can be defined with the sample average $g^2=\langle(d'_{hkl})^2\rangle/d_{hkl}^2-1$.[18, 19] For real materials, $g$ varies from 0-15%, where <1% is indicative of highly crystalline behavior, 1-10% represents paracrystalline materials, and 10-15% is considered to be a glass or melt.[16] Hence $g$ can be used to rank materials quantitatively from perfectly ordered to completely disordered on a continuous scale.

In polymer crystallites, π-π stacking gives rise to substantial wavefunction overlap allowing 2D delocalization of charges and excitations.[20, 21] Thus, carrier mobility depends on the intermolecular charge transfer rate in the π-π stacking direction. Therefore, structural disorder in this direction must strongly affect transport. There are many ways in which disorder can occur, such as conformational changes (molecular tilts and twists), defects (stacking faults and dislocations), or positional fluctuations in various crystallographic directions. Here we focus on disorder caused by static positional fluctuations. We measure quantitatively the $g$ parameter in the crystallites of a high-performance semiconducting polymer. We hypothesize that lattice spacing modulations in the crystallites along the π-π stacking direction lead to trap states in the bandgap of semicrystalline semiconducting polymers. We then use first-principles calculations to show that realistic values of *g* generate trap states in the bandgap of this polymer that are consistent with experimental measurements.

In order to study the effect of disorder within the crystallites on band structure and transport, we choose the fused-ring polythiophene PBTTT, poly[2,5-bis(3-alkylthiophen-2-yl)thieno(3,2-b)thiophene], as a model material. With reported hole mobilities of 1 cm$^2$/Vs, PBTTT is one of the best performing polymeric semiconductors to date.[22, 23] This material exhibits outstanding order when deposited in thin films and annealed, as demonstrated in numerous studies.[23, 24] In thin films, which have a strong fiber texture, the two directions associated with charge transport (the conjugated backbone, and the close π-π stacking directions) are found in the plane of the substrate and form 2D sheets.[21, 24, 25] Crystallites in PBTTT are characterized by liquid-crystalline-like packing arrangements. It is within these ordered regions that we wish to measure paracrystallinity.

Paracrystalline disorder can be measured experimentally using advanced X-ray lineshape analysis to quantitatively extract the fluctuations in molecular spacing due specifically to cumulative disorder in the π-π stacking direction. Scattering measurements were performed using high-resolution grazing incidence X-ray diffraction at the Stanford Synchrotron Radiation Lightsource, with a photon energy of 8 keV, and an incident angle of 0.25º. All diffraction experiments were carried out with the samples enclosed in a Helium-filled chamber in order to reduce the effects of air scattering and beam damage due to the intense X-ray beam.

For the analysis of the diffraction lineshapes we use the Fourier transform formalism developed by Warren and Averbach,[19] and extended by others.[26, 27] In this formalism, two main sources of broadening are included: a diffraction-order-independent component that reflects the size of the crystallites, and an order-dependent portion that is affected by paracrystalline displacements as well as variations in the average lattice spacing. In this model, each diffraction peak is constructed from the superposition of waves scattered by unit cells whose positional distortions from the ideal lattice are described by Gaussian statistics and belong to columns of cells along the [hkl] direction. The normalized *n*-th coefficient of the Fourier transform of the *m*-th order diffraction peak is then be given by

$$A_m(n) = A_m^S(n) \exp\left[-2\pi^2 m^2 \left(n g^2 + n^2 e_{RMS}^2\right)\right],$$

where $A^S_m(n)$ is the size-related broadening contribution which depends on the column length distribution in the sample (related to crystallite size), $e_{RMS}$ is the lattice parameter fluctuation reduced variance,[18] and the rest of the variables are as previously defined. It must be noted that the XRD intensity as a function of the scattering vector, *q*, is the reciprocal space mapping of the real space crystal lattice, and thus its Fourier transform represents the coherence of the material in real space, with the coefficients for large *n* describing the correlation between units located a distance $nd_{hkl}$ apart. The details of this analysis are treated elsewhere.[18, 19, 28]

Because this analysis requires data from multiple diffraction orders, it is especially challenging to apply it to diffraction from the π-π stacking planes of spin-cast polymers, where the isotropic-in-plane crystallite orientation spreads the diffracted intensity over the entire azimuthal angular range. To overcome this difficulty, films of PBTTT were deposited via a previously described flow coating technique, and heated to form films of aligned ribbon phase.[29] In this phase, the molecules are aligned with their chain backbones parallel to the flow coating direction, and the π-π stacking along the in-plane direction perpendicular to the flow. This alignment allows for the decoupling of the structural information specific to each direction.[30-32] Additionally, by aligning the film, diffraction from a particular crystallographic direction is concentrated into a narrower region of reciprocal space, providing higher intensities which allow a higher signal to noise ratio and the detection of multiple-order reflections (010, 020, …).

The grazing incidence diffraction pattern associated with the π-π stacking direction of an aligned PBTTT film, denoted as the [010] direction,[24] is shown in Figure 1a. Clearly visible are the first two diffraction orders and a weak third order of the (0k0) family. Also evident are contributions from the chain backbone peaks, indicative of imperfect macroscopic alignment of the polymer chains. By fitting the full diffraction profile, accounting for all peaks and background scattering, we are able to appropriately isolate the (0k0) peaks. The Fourier transforms of the first two orders of the π-π stacking direction (Fig. 1b) are then fit to the model described above. Uncertainties in the measured diffracted intensity are assessed and propagated through the entire analysis, allowing for conservative estimates of confidence bounds for the calculated fitting results. The inclusion of crystallite size effects in the model adds no information to the fits as evidenced by an invariance of the fitting results with respect to the parameters describing different grain size distribution models.[28] In the limit where the effect of crystallite size is negligible, $A^S_m(n)=1$, the only parameters describing the Fourier transform of the lineshape are $g$ and $e_{RMS}$, and the fitted values are negligibly different from those fitted including crystallite size effects. This observation agrees with recent findings that in-plane order in PBTTT has a liquid-crystalline-like nature, where there are gradual transitions from one orientation to another,[33] making it difficult to define a grain size. We find that $g$=7.3%±0.2%, and $e_{RMS}$=0.9%±0.1%. It is again informative to remember that a material with a paracrystalline parameter of 10% is considered to be amorphous. Hence, despite the well defined, regular morphology of these aligned films, and the extremely long range coherence along the lamellar stacks, the π-π stacking direction shows an undeniably disordered behavior. In the context of charge transport, the direction with the highest molecular orbital overlap will influence the electronic properties of the material most strongly. For this reason, quantifying positional fluctuations along the π-π stacking and chain backbone repeats is most relevant for understanding the effects of structural order on transport.[34] On the other hand, there may be little connection between transport and lamellar stacking quality.

First-principles calculations are employed to relate structural disorder to electronic disorder. Calculations of the position of the valence band maximum (VBM) as a function of the π-π stacking distance provide evidence that disorder in this direction can give rise to states having energies within the band gap of the perfect crystal. Previous studies on polythiophenes show that the energy of the VBM increases as the π-π stacking distance is reduced,[6] and deviations of the angle of tilt[25] of the polythiophene backbone give rise to states in the gap.[35] In order to construct a simple model Hamiltonian to describe this system, density functional theory calculations were performed using previously discussed techniques.[36]

Calculations for systems having four PBTTT chains per unit cell with an alternating expansion and compression of the π-π stacking distance were performed. The length of the unit cell in the π-π stacking direction was taken to be $4a$, and the separations between the four PBTTT chains were ($a+\Delta a$, $a-\Delta a$, $a-\Delta a$, $a+\Delta a$), with $a$=3.8 Å and for distortions ranging from $\Delta a$=0 to $\Delta a$=0.51 Å. These distortions introduce states with energies within the bandgap of the perfect crystal, changing the position of the VBM for the distorted system (Fig. 2).

We employ these results to construct a simple model Hamiltonian for a collection of polymer chains with one orbital per site to model the dispersion of the highest occupied molecular orbital (HOMO) along the polymer backbone and in the π-π stacking direction. This allows us to calculate the distribution of gap states for a two-dimensional disordered system. Neighboring sites along the polymer chain are connected by a constant matrix element $h_o$, and neighboring sites along the π-π stacking direction are connected by a separation-dependent matrix element $t$. We neglect the possibility of disorder in the on-site energy. In order to reproduce the energy dispersion of the HOMO along the polymer backbone, we take $h_o$=450 meV.[36] The magnitude and distance dependence of $t$ is obtained from the first principles calculations on the reduced system described above. With an exponential dependence of $t$ on $\Delta a$, $t = t_0 \exp(-\beta \cdot \Delta a)$, the model Hamiltonian is able to reproduce the dependence of the VBM energy on $\Delta a$ obtained in the first principles calculations. We find $t_0$=150 meV and $\beta$=2.35 Å$^{-1}$. A similar set of parameters has been obtained for P3HT.

Consider a crystallite comprised of 50 π-π stacked chains, with 20 sites along the backbone in each chain. This results in a Hamiltonian matrix $H$ of order 1000×1000. The off-diagonal elements $t$ are determined by choosing interchain distances from a Gaussian distribution having a standard deviation $\sigma$, which corresponds to a paracrystallinity $g=\sigma/a$. For each matrix $H$ we solve the equation $H\psi_n=E_n\psi_n$ to obtain the energy eigenvalues $E_n$ and eigenfunctions $\psi_n$. For each value of $g$ we average over an ensemble consisting of $10^4$ crystallites in order to determine the DOS in the band tail region. This produces a sufficient number of states to compute the band tail distribution over several orders of magnitude. The resulting DOS, $N(E)$, is shown for various values of $g$ in Figure 2. In spite of the Gaussian disorder, this model predicts an exponential dependence of the band tail distribution, with an energy spread that increases with the amount of paracrystalline disorder. One can estimate the slope of the exponential band tail $E_0$ by fitting the density of states in the band tail region with the function $N(E)=N_0 \exp(-E/E_0)$, with an increasing energetic breadth of the tail (Fig.2) with values near 10 meV for highly ordered systems and 100 meV for strongly disordered ones. Examination of the eigenvectors $\psi_n$ indicates that the states with energies lying within the bandgap of the unperturbed system are localized on just a few polymer chains. States with energies well below the VBM band edge tend to be less localized. The total number of states in each calculation is constant, showing a greater number of localized tail states as the amount of disorder increases.

In conclusion, advanced X-ray lineshape analysis of PBTTT indicates the presence of a large amount of paracrystalline disorder along the π-π stacking direction ($g$=7.3%) and modeling of such a system results in exponential tails in the DOS of the crystallites with a breadth $E_0$~100 meV. By modeling the band structure of a collection of π-π stacked PBTTT segments with increasing amounts of paracrystalline disorder it is possible to show that, compared to an ideally ordered microstructure, the experimentally determined amount of disorder introduces a tail of localized states which can act as traps for charge transport with an energetic distribution consistent with experimental results.[10, 11] These calculations provide physical justification for the mobility edge model.[10, 37]

*Acknowledgments:*
*Portions of this research were carried out at the Stanford Synchrotron Radiation Lightsource, a national user facility operated by Stanford University on behalf of the U.S. Department of Energy, Office of Basic Energy Sciences. A.S. and J.R. gratefully acknowledge financial support from the National Science Foundation in the form of, respectively, a Career Award and a Graduate Student Fellowship. This publication was partially based on work supported by the Center for Advanced Molecular Photovoltaics (Award No KUS-C1-015-21), made by King Abdullah University of Science and Technology (KAUST). Work at the Palo Alto Research Center (PARC) was supported by the AFOSR under Grant No. FA9550-09-1-0436.*

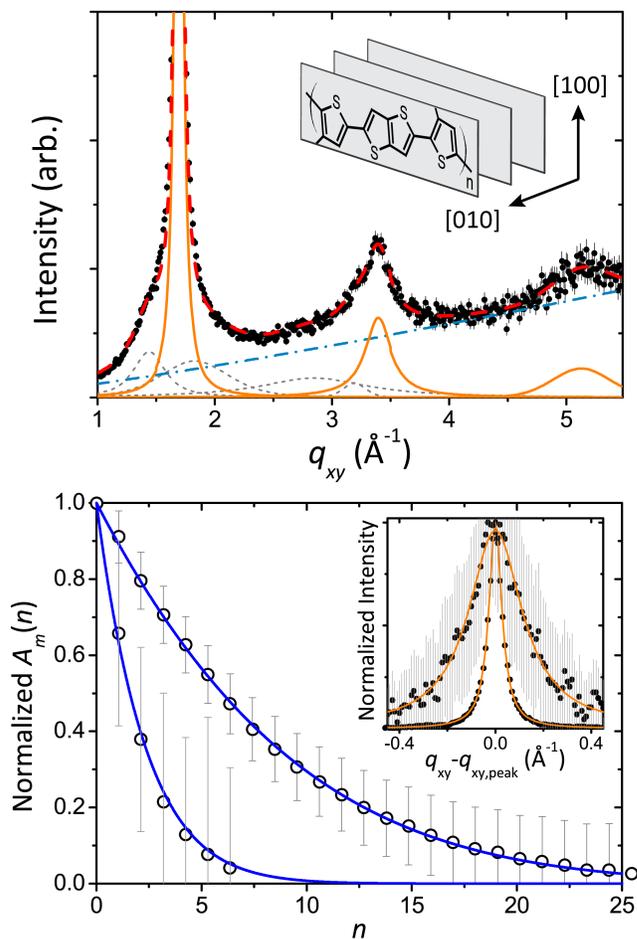

Figure 1. Top: Grazing Incidence X-ray scattering of aligned ribbon phase PBTTT in the π-π stacking direction (circles). Fits used to isolate the π-π stacking peaks (0k0): the peaks in question are shown (solid lines), chain backbone related peaks (dotted lines), background (dash-dotted line), and the complete fit (dashed line). Inset: the general packing motif of PBTTT, with the π-π stacking and lamellar stacking direction indicated. Bottom: The Fourier transforms of the isolated (010) and (020) peaks with the fit to the Warren-Averbach model. Inset: normalized (0k0) peaks showing order-dependent broadening.

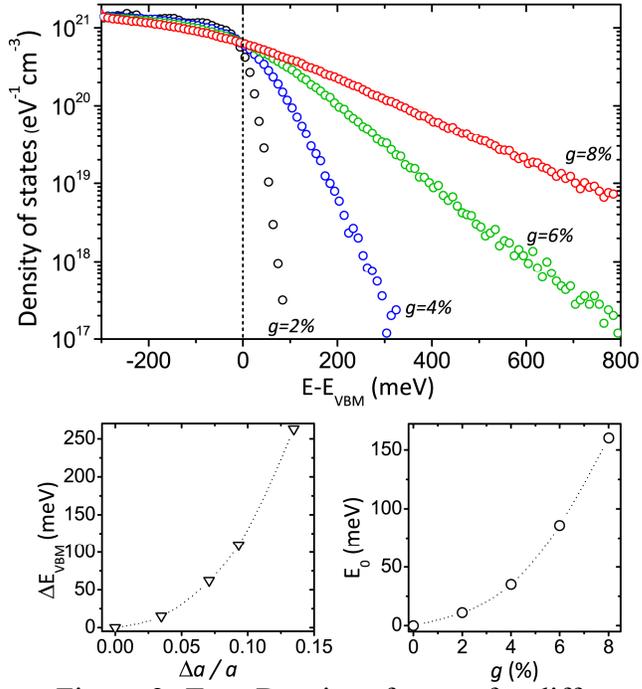

Figure 2. Top: Density of states for different values of paracrystalline disorder. The energy $E_{VBM}$ corresponds to the valence band maximum in a perfectly ordered crystalline region. Bottom left: Change in the position of the valence band maximum for a distorted PBTTT crystal. Bottom right: Slope of the exponential tails of the DOS shown in the top panel.

References:

[1] H. Klauk, *Organic Electronics: Materials, Manufacturing and Applications* (Wiley-VCH, Weinhiem Germany, 2006).
[2] A. Salleo, Materials Today **10**, 38 (2007).
[3] V. Podzorov, E. Menard, J. A. Rogers, and M. E. Gershenson, Physical Review Letters **95**, 226601 (2005).
[4] S. Baranovskii, *Charge Transport in Disordered Solids with Applications in Electronics* (Wiley, 2006).
[5] D. P. McMahon and A. Troisi, ChemPhysChem **11**, 2067 (2010).
[6] R. A. Street, J. E. Northrup, and A. Salleo, Physical Review B **71**, 165202 (2005).
[7] T.-J. King, M. G. Hack, and I.-W. Wu, Journal of Applied Physics **75**, 908 (1994).
[8] G. Horowitz, R. Hajlaoui, D. Fichou, and A. El Kassmi, Journal of Applied Physics **85**, 3202 (1999).
[9] W. L. Kalb and B. Batlogg, Physical Review B **81**, 035327 (2010).
[10] A. Salleo, T. W. Chen, A. R. Volkel, Y. Wu, P. Liu, B. S. Ong, and R. A. Street, Physical Review B **70**, 115311 (2004).
[11] C. Wang, L. H. Jimison, L. Goris, I. McCulloch, M. Heeney, A. Ziegler, and A. Salleo, Advanced Materials **22**, 697 (2009).
[12] N. Zhao, Y. Y. Noh, J. F. Chang, M. Heeney, I. McCulloch, and H. Sirringhaus, Advanced Materials **21**, 3759 (2009).
[13] J. Klafter and J. Jortner, Chemical Physics **26**, 421 (1977).
[14] J. H. Kang, D. d. S. Filho, J.-L. Bredas, and X.-Y. Zhu, Applied Physics Letters **86**, 152115 (2005).
[15] G. Strobl, *The Physics of Polymers* (Springer, New York, 1997).
[16] A. M. Hindeleh and R. Hosemann, Journal of Materials Science **26**, 5127 (1991).
[17] R. Bonart, R. Hosemann, and R. L. McCullough, Polymer **4**, 199 (1963).
[18] T. J. Prosa, J. Moulton, A. J. Heeger, and M. J. Winokur, Macromolecules **32**, 4000 (1999).
[19] B. E. Warren, *X-Ray Diffraction* (Addison-Wesley, Reading, MA, 1969).
[20] X. M. Jiang, R. Österbacka, O. Korovyanko, C. P. An, B. Horovitz, R. A. J. Janssen, and Z. V. Vardeny, Advanced Functional Materials **12**, 587 (2002).
[21] H. Sirringhaus, et al., Nature **401**, 685 (1999).
[22] B. H. Hamadani, D. J. Gundlach, I. McCulloch, and M. Heeney, Applied Physics Letters **91**, 243512 (2007).
[23] I. McCulloch, et al., Nature Materials **5**, 328 (2006).
[24] For notational simplicity, we use the indexing scheme of M. L. Chabinyc, M. F. Toney, R. J. Kline, I. McCulloch, and M. Heeney, Journal of the American Chemical Society **129**, 3226 (2007).
[25] D. M. DeLongchamp, R. J. Kline, E. K. Lin, D. A. Fischer, L. J. Richter, L. A. Lucas, M. Heeney, I. McCulloch, and J. E. Northrup, Advanced Materials **19**, 833 (2007).
[26] B. Crist and J. B. Cohen, Journal of Polymer Science: Polymer Physics Edition **17**, 1001 (1979).
[27] H. Takahashi, Journal of the Physical Society of Japan **27**, 708 (1969).
[28] J. Rivnay, R. Noriega, L. H. Jimison, R. J. Kline, M. F. Toney, and A. Salleo, In preparation.
[29] D. M. DeLongchamp, et al., ACS Nano **3**, 780 (2009).
[30] J. Rivnay, M. F. Toney, Y. Zheng, I. V. Kauvar, Z. Chen, V. Wagner, A. Facchetti, and A. Salleo, Advanced Materials **22** (2010).
[31] H. N. Tsao, D. Cho, J. W. Andreasen, A. Rouhanipour, D. W. Breiby, W. Pisula, and K. Müllen, Advanced Materials **21**, 209 (2009).
[32] D. M. DeLongchamp, R. J. Kline, D. A. Fischer, L. J. Richter, and M. F. Toney, Advanced Materials, n/a (2010).
[33] X. Zhang, S. D. Hudson, D. M. DeLongchamp, D. J. Gundlach, M. Heeney, and I. McCulloch, Advanced Functional Materials **20** (2010).
[34] J. L. Bredas, J. P. Calbert, D. A. da Silva Filho, and J. Cornil, Proceedings of the National Academy of Sciences of the United States of America **99**, 5804 (2002).
[35] J. E. Northrup, Unpublished work (2010).
[36] J. E. Northrup, Physical Review B **76**, 245202 (2007).
[37] J.-F. Chang, H. Sirringhaus, M. Giles, M. Heeney, and I. McCulloch, Physical Review B **76**, 205204 (2007).